\documentclass[aps,pra,twocolumn,amsmath,amssymb,superscriptaddress,longbibliography]{revtex4-1}
\usepackage[english]{babel}

\usepackage{amssymb}
\usepackage{dcolumn}
\usepackage{bm}
\usepackage{graphicx}
\usepackage{amsmath}

\usepackage{graphicx}        
\graphicspath{{pict/}{}}

\usepackage{dcolumn}
\usepackage{bm}
\usepackage[dvipdfm, pdfstartview=FitH, CJKbookmarks=true, bookmarksnumbered=true, bookmarksopen=true, colorlinks=true, pdfborder=001, citecolor=blue, linkcolor=blue, urlcolor=blue, linktocpage=true] {hyperref}

\begin{document}

\title{Symmetry-protected topological phase for spin-tensor-momentum-coupled ultracold atoms}

\author{Zhoutao Lei}
\affiliation{Guangdong Provincial Key Laboratory of Quantum Metrology and Sensing $\&$ School of Physics and Astronomy, Sun Yat-Sen University (Zhuhai Campus), Zhuhai 519082, China}
\affiliation{State Key Laboratory of Optoelectronic Materials and Technologies, Sun Yat-Sen University (Guangzhou Campus), Guangzhou 510275, China}

\author{Yuangang Deng}
\email{dengyg3@mail.sysu.edu.cn}
\affiliation{Guangdong Provincial Key Laboratory of Quantum Metrology and Sensing $\&$ School of Physics and Astronomy, Sun Yat-Sen University (Zhuhai Campus), Zhuhai 519082, China}

\author{Chaohong Lee}
\email{lichaoh2@mail.sysu.edu.cn}
\affiliation{Guangdong Provincial Key Laboratory of Quantum Metrology and Sensing $\&$ School of Physics and Astronomy, Sun Yat-Sen University (Zhuhai Campus), Zhuhai 519082, China}
\affiliation{State Key Laboratory of Optoelectronic Materials and Technologies, Sun Yat-Sen University (Guangzhou Campus), Guangzhou 510275, China}
\affiliation{Synergetic Innovation Center for Quantum Effects and Applications, Hunan Normal University, Changsha 410081, China}

\date{\today}

\begin{abstract}
We propose a realizable experiment scheme to construct a one-dimensional synthetic magnetic flux lattice with spin-tensor-momentum coupled spin-1 atoms and explore its exotic topological states.
Different from the Altland-Zirnbauer classification, we show that our system hosts a symmetry-protected phase protected by a magnetic group symmetry (${\cal M}$) and characterized by a $Z_2$ topological invariant.
In single-particle spectra, we show that the topological nontrivial phase supports two kinds of edge states, which include two (four) zero-energy edge modes in the absence (presence) of two-photon detuning.
We further study the bulk-edge correspondence in a non-Hermitian model by taking into account the particle dissipation.
It is shown that the non-Hermitian system preserves the bulk-edge correspondence under the ${\cal M}$ symmetry but exhibits the non-Hermitian skin effect with breaking the ${\cal M}$ symmetry at nonzero magnetic flux.
This work provides insights in understanding the exotic topological quantum states of high-spin systems and facilitating their experimental explorations.
\end{abstract}

\maketitle

\section{introduction\label{Sec1}}
Topological quantum matters, which are characterized by gapped bulk states and symmetry-protected gapless edge states, drive much of the fundamental research ranging from condensed-matter physics~\cite{RevModPhys.82.1959,RMPKane2010QSHE,RMPXL2011SOC,RevModPhys.88.035005} to quantum information processing~\cite{PhysRevLett.105.077001,RevModPhys.80.1083}.
The paradigms of topological states include integer quantum Hall effects~\cite{PRLKlitzing1980IQHE,PRLThouless1982IQHE,AnnKohmoto1985IQHE}, quantum spin-Hall effects~\cite{PRLKane2005QSHE1,PRLKane2005QSHE2,PRLZhang2006QSHE,Bernevig1757,Roth294}, Majorana fermions~\cite{Mourik1003,MLDMT2012,NPDA2012}, and Weyl semimetals~\cite{Xu613,Lu622,PhysRevX.5.031013,NSA2015,PhysRevX.9.021053}.
Besides the topologically ordered phases with long-range quantum entanglement, symmetry-protected topological (SPT) phases whose edge states are only robust against local perturbations have attracted much attention in recent years~\cite{PhysRevB.83.035107,PhysRevB.84.235128,SCIWXG2012}.
However, the realization of SPT phases remains a challenging task, despite many classes of SPT phases that have been predicted in various theoretical proposals~\cite{PhysRevLett.50.1153,PhysRevB.86.115109,PhysRevB.87.155114,PhysRevLett.119.246402,PhysRevLett.122.180401} and few of them are realized in solid-state materials~\cite{RevModPhys.88.035005}.

Meanwhile, the recent experimental breakthroughs of spin-orbit (SO) coupling in ultracold atoms~\cite{NATYJ2011SOCAT, PRLZW2012SOCAT, SCPJW2016SOC, NSHL2016} have provided a new paradigm for exploring a variety of topological states in a clean environment and controllable way~\cite{RevModPhys.83.1523, Zhai_2015, Goldman_2014, RevModPhys.91.015005}.
In particular, the successful realization of Raman-assisted tunneling in optical lattices~\cite{PhysRevLett.107.255301,NPKCJ2015} could realize the strong synthetic magnetic fields and establish the tunable magnetic flux lattices~\cite{PhysRevLett.103.035301,PhysRevLett.105.190404,PhysRevLett.105.255302,PhysRevLett.111.185301,PhysRevLett.111.185302,NaJG2014,NPAM2014}. Up to now, the realized SO couplings and the observed SPT phases are mainly focusing on the spin-$1/2$ quantum gases~\cite{SCiGB2018TOPO,NSBS2019,scileuc775}.
Unlike the Rashba spin-vector-momentum coupling, a variety of spin-tensor-momentum coupling (STMC) could be constructed for high-spin system, which could provide exotic quantum  phenomena~\cite{PRLCZ2017STMC,PRLCZ2018STMC,PhysRevA.97.053609}. Moreover, due to their particle gain and loss, non-Hermitian systems could host striking quantum phenomena and applications, giving rise to the exceptional points~\cite{Zhen2015, Xu2016, SCIHZ2018EP, Mirieaar7709, Ozdemir2019}, parity-time ($\mathcal{P}\mathcal{T}$) symmetry breaking transitions ~\cite{PhysRevLett.110.243902,PhysRevB.84.153101, PhysRevLett.115.200402, NatmatWei2017,Xiao2017}, topological insulator lasers~\cite{PhysRevLett.120.113901,Hararieaar4003,Bandreseaar4005}, and quantum brachistochrone problem~\cite{PhysRevLett.98.040403, PhysRevLett.99.130502, Bender_2007, PhysRevLett.101.230404, PhysRevLett.110.050403}.
Remarkably, the bulk spectra in the non-Hermitian systems qualitatively depend on the boundary condition (non-Hermitian skin effect)~\cite{PRLTE2016AE,JPCXY2018NS,PRLFK2018NT,PRLYS12018NT,PRLYS212018NT, PhysRevX.8.031079,PRBSZ2019NS,PhysRevLett.123.066404,PhysRevLett.123.170401,PhysRevLett.123.246801}, which is beyond the paradigmatic bulk-edge correspondence in Hermitian system~\cite{RevModPhys.82.1959,RMPKane2010QSHE,RMPXL2011SOC,RevModPhys.88.035005}.
It is of great interest to explore \emph{whether new SPT phases can emerge via STMC in both Hermitian and non-Hermitian systems of ultracold atoms}.
An affirmative answer will significantly facilitate the experimental explorations of exotic topological quantum matters in high-spin physics~\cite{PRAZHU2017TOPO, PRAZHU2018STMC, PhysRevLett.120.130503}.

In this work, we show how to realize STMC using Raman-assisted staggered spin-flip hoppings in pseudospin-1 ultracold atoms and then study its topological states.
Due to the interplay of synthetic magnetic flux and two-photon detuning, the system exhibits a SPT phase which is beyond the Altland-Zirnbauer (AZ) classification.
This SPT phase satisfies a magnetic group symmetry (${\cal M}$) and supports two kinds of edge states.
Without loss of generality, we further study the corresponding non-Hermitian models with in-plane (${\cal B}_x$) and out-of-plane (${\cal B}_z$) imaginary magnetic fields, respectively.
Strikingly, the ${\cal M}$ symmetry can guarantee the bulk-edge correspondence even in the presence of nonzero ${\cal B}_x$ field.
Our result is different from the non-Hermitian spin-$1/2$ models~\cite{JPCXY2018NS, PRLFK2018NT, PRLYS12018NT, PRLYS212018NT, PhysRevX.8.031079, PRBSZ2019NS, PhysRevLett.123.066404, PhysRevLett.123.170401, PhysRevLett.123.246801}, where the bulk-edge correspondence is broken when the ${\cal B}_x$ field is applied.
Furthermore, the non-Hermitian skin effects in which the bulk eigenstates are localized near the boundary is predicted ascribe to the ${\cal M}$ symmetry breaking when ${\cal B}_z$ field applied at nonzero magnetic flux.

This paper is organized as follow. In Sec~\ref{Sec2}, we introduce our model of the Raman-assisted STMC and derive the system Hamiltonian. In Sec~\ref{Sec3}, we study the topological states and analyze the property of edge states. In Sec~\ref{Sec4}, we present the bulk-edge correspondence in non-Hermitian models. Finally, a brief summary is given in Sec~\ref{Sec5}.

\section{model and hamiltonian\label{Sec2}}
\begin{figure}[!htp]
\includegraphics[width=0.95\columnwidth]{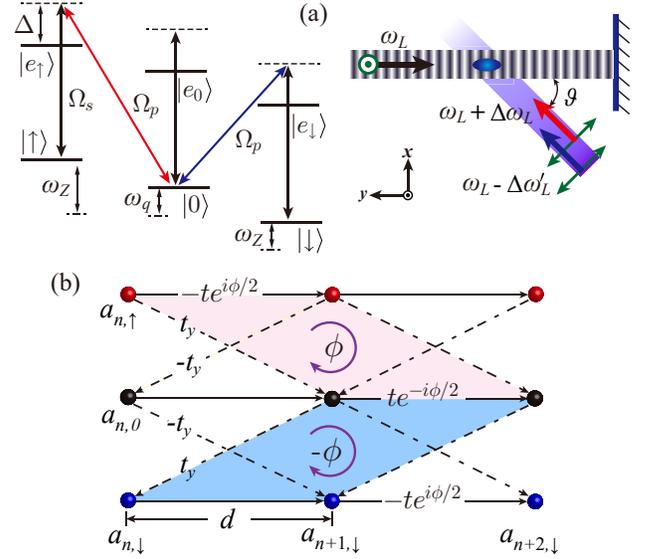}
\caption{\label{model}(color online). (a) Level diagram and schematic for creating STMC for ultracold atoms. (b) Proposed scheme for generating synthetic magnetic-flux $\phi$ ($-\phi$) per upper (lower) half plaquette via combining spatially dependent complex hopping with Raman-assisted hopping.}
\end{figure}

We consider a quantum gas of noninteracting ultracold atoms subjected to a bias magnetic field ${\bf B}$ along the quantization $z$ axis.
The three atomic ground states form a pseudospin-$1$ manifold and their corresponding linear (quadratic) Zeeman shifts are $\hbar\omega_Z$ ($\hbar \omega_q$).
In Fig.~\ref{model}(a), we illustrate the atomic level structure and laser configuration.
The atomic transition $|\sigma\rangle\leftrightarrow |e_{\sigma}\rangle$ is coupled by a $\pi$-polarized standing-wave laser with frequency $\omega_L$ and Rabi frequency $\Omega_s(y)=\Omega_s\cos(k_L y)$, where $k_L$ is the wave vector and $\sigma=\{\uparrow,\downarrow,0\}$.
To achieve Raman transitions, the atomic transition $|0\rangle\leftrightarrow |e_{\uparrow}\rangle$ ($|0\rangle\leftrightarrow |e_{\downarrow}\rangle$) is driven by $\sigma$-ploarized plane-wave lasers with frequencies $\omega_L+\Delta \omega_L$ ($\omega_L-\Delta \omega'_L$) for matching the Zeeman shifts and Raman selection rules.
Here the Rabi frequency for the two plane-wave lasers is given as $\Omega_p(y)=\Omega_p e^{-i\kappa y}$ with $\kappa=k_L\cos\vartheta$.
We should emphasize that the off-resonant Raman processes are suppressed for sufficiently large quadratic Zeeman shifts~\cite{NATYJ2011SOCAT}, where the linear Zeeman shift is compensated by the frequency difference of the Raman fields.
For a large light-atom detuning, i.e. $|\Omega_{s,p}/\Delta|\ll 1$ and $\Delta=\omega_L-\omega_a$, one can adiabatically eliminate the excited states $|e_{\sigma}\rangle$.
Therefore the light-atom interaction results in a one-dimensional spin-independent optical lattice ${\cal U}_{\rm ol}(y)=U_s\cos^2(k_Ly)\hat{I}$ with Stark shift $U_s=-\Omega_s^2/\Delta$ and lattice constant $d=\pi/k_L$.
By applying an analogous gauge transform \cite{PRBYD2017TOPO},
\begin{eqnarray*}
\left|\uparrow\right\rangle \rightarrow e^{i\kappa y/2}\left|\uparrow\right\rangle, \left|\downarrow\right\rangle~\rightarrow e^{i\kappa y/2}\left|\downarrow\right\rangle, \left|0\right\rangle \rightarrow e^{-i\kappa y/2}\left|0\right\rangle,
\label{Eq.transform}
\end{eqnarray*}
the single-particle Hamiltonian reads
\begin{eqnarray}
{\boldsymbol{h}}_0=\frac{({\mathbf{p}}-{\mathbf{A}})^{2}}{2M}+\Omega\cos(k_Ly) \hat{F}_{x}+{\delta}\hat{F}_{z}+{\cal U}_{\rm{ol}}(x)\hat{I}, \label{Eq.Ham 0}
\end{eqnarray}
where $M$ is the atomic mass, $\hat{I}$ is the identity matrix, $\hat{F}_{x,y,z}$ is the spin-1 matrix, $\Omega=-\sqrt{2}\Omega_s\Omega_p/\Delta$ is the Raman coupling strength, and $\delta=\omega_Z+2\Omega_p^2/\Delta-\Delta \omega_L-\omega_q$ is the effective two-photon detuning under the condition of $\Delta \omega_L'=\Delta \omega_L-2(\omega_q-2\Omega_p^2/\Delta)$.
In particular, ${\mathbf{A}}=-\hbar\kappa[2\hat{F}^2_{z}-\hat{I}]/2$ is the vector potential, which denotes the STMC with the SO coupling strength $\kappa$.

For a sufficiently strong lattice potential with blue detunings (i.e. $U_s>0$), the tight-banding Hamiltonian with considering the lowest orbit and the nearest-neighbor hoppings takes the form
\begin{eqnarray}\label{Eq.Ham 1}
{H_{0}}&\!=\!&\sum_{\sigma=\uparrow,\downarrow}\sum_{n}\bigg[(-1)^{n}t_{y} (\hat{a}^{\dag}_{n,\sigma}\hat{a}_{n+1,0} -\hat{a}^{\dag}_{n,\sigma}\hat{a}_{n-1,0}) \nonumber \\
&-&t(\hat{a}^{\dag}_{n,\sigma} \hat{a}_{n+1,\sigma}e^{-i\phi/2} +  \hat{a}^{\dag}_{n,0}\hat{a}_{n+1,0}e^{i\phi/2})  +{\rm H.c.})\bigg]\nonumber \\
&+&{\delta}\sum_{n}(\hat{a}^{\dag}_{n,\uparrow}\hat{a}_{n,\uparrow}-\hat{a}^{\dag}_{n,\downarrow}\hat{a}_{n,\downarrow}),
\end{eqnarray}
where $\hat{a}_{n,\sigma=\uparrow,\downarrow,0}$ is the atomic annihilation operator for the $n$th site, $t$ is the nearest-neighbor spin-independent hopping, $t_y$ is the Raman-assisted  nearest-neighbor spin-flip hopping, and $\phi=\kappa d$ is the Peierls phase.
With our laser configuration shown in Fig.~\ref{model}(a), the synthetic magnetic flux ($-1<\phi/\pi<1$) can be easily tuned by changing the angle $\vartheta$ with respect to the $y$ axis. In addition, the on-site spin-flip hopping is zero with netting the nearest-neighbor spin-flip hopping and corresponding the atoms symmetrically localized at the nodes for the blue lattice potential~\cite{PRBYD2017TOPO}.

To gain more insights into the magnetic flux, we introduce the gauge transformation $\hat{a}^{\dag}_{n,0} \longrightarrow
(-1)^{n+1}\hat{a}^{\dag}_{n,0}$ to eliminate the staggering factor in the spin-flip hopping.
Then the lattice Hamiltonian becomes
\begin{eqnarray}\label{Eq.Ham 2}
{H_{0}}&\!=\!&\sum_{\sigma=\uparrow,\downarrow}\sum_{n}\bigg[ t_{y} (\hat{a}^{\dag}_{n,\sigma}\hat{a}_{n+1,0} -\hat{a}^{\dag}_{n,\sigma}\hat{a}_{n-1,0}) \nonumber \\
&-&t(\hat{a}^{\dag}_{n,\sigma} \hat{a}_{n+1,\sigma}e^{-i\phi/2} -  \hat{a}^{\dag}_{n,0}\hat{a}_{n+1,0}e^{i\phi/2})  +{\rm H.c.})\bigg]\nonumber \\
&+&{\delta}\sum_{n}(\hat{a}^{\dag}_{n,\uparrow}\hat{a}_{n,\uparrow}-\hat{a}^{\dag}_{n,\downarrow}\hat{a}_{n,\downarrow}),
\end{eqnarray}
whose corresponding schematic is shown in Fig.~\ref{model}(b).
In contrast to nonzero flux for the spin-half model~\cite{PRBYD2017TOPO}, the net synthetic magnetic flux per plaquette is zero due to the STMC for our spin-$1$ system.

Under the periodic boundary condition (PBC) satisfying translational invariance, the Hamiltonian in the momentum space is given as
\begin{align}
{\boldsymbol{h}}_0({k}) = \epsilon({ k})\hat{I} +\delta\hat{F_{z}}+d_{1}({ k})\hat{Q}_{yz} +d_{2}({k})\hat{F}_{z}^2,
 \label{k-single}
\end{align}
where $\epsilon({ k}) =2t\cos(kd+\phi/2)$, $d_{1}({ k})=-2\sqrt{2}t_{y}\sin(kd)$, $d_{2}({ k})=-4t\cos(kd)\cos(\phi/2)$, $\hat{F}_{z}^2$ is the spin-quadrupolar operator, and $\hat{Q}_{yz}$ is the generator of SU(3) Lie algebra~\cite{PRA88033629}. Interestingly, the last two terms of $\hat{Q}_{yz}$ and $\hat{F}_{z}^2$ in Eq. (\ref{k-single}) represent two different types of STMC, which play essential roles for forming topological states.

To explore the symmetry classes, we introduce three symmetry operators for spin-$1$ systems:
the time reversal symmetry ${\cal T}=e^{-i\pi \hat{F}_y}K$, the particle-hole symmetry ${\cal C}=ie^{-i\pi\hat{F}_x}K$, and the chiral symmetry ${\cal S}={\cal C}\otimes {\cal T}=ie^{-i\pi\hat{F}_z}$ (defined as the product of ${\cal T}$ and ${\cal C}$), where $K$ is a complex conjugate operator.
Explicitly, the gauge-potential-induced dispersion $\epsilon({ k})$ simultaneously breaks ${\cal T}$, ${\cal C}$, and ${\cal S}$ symmetries for nonzero magnetic flux.
Due to the STMC, the term of $d_{1}({ k})\hat{Q}_{yz}$ breaks both ${\cal T}$ and ${\cal C}$ symmetries, while the term of $d_{2}({k})\hat{F}_{z}^2$ breaks both ${\cal C}$ and ${\cal S}$ symmetries.
Therefore, the Hamiltonian~(\ref{k-single}) is beyond the conventional AZ classification for the one-dimensional (1D) system in the absence of the particle-hold or chiral symmetry protection~\cite{PRBAA1997TOCL}.
However, we find that the Hamiltonian ${\boldsymbol{h}}_0({k})$ satisfies a magnetic group symmetry ${\cal M}{\boldsymbol{h}}_0({k}){\cal M}^{-1}={\boldsymbol{h}}_0({-k})$ with ${\cal M}=e^{-i \pi\hat{F}_z}{\cal K}\otimes {\cal R}_y$, which is a combination of ${\cal T}$ and the mirror symmetry $M_y=e^{-i\pi \hat{F}_{x}}R_y$ with $R_y$ representing the spatial reflection along the $y$ axis.
In addition, the introduced magnetic group symmetry satisfies $[{\cal M},{\boldsymbol{h}}_0({k})]=0$ and ${\cal M}={\cal M}^{-1}$ so that it brings a SPT phase hosting a topological nontrivial phase characterized by the 1D $Z_2$ invariant~\cite{Ryu_2010,SCiGB2018TOPO}.

\section{Band topology and edge states\label{Sec3}}
To further characterize the SPT phase, we calculate the energy spectrum in $k$ space via ${\boldsymbol{h}}_0({k})|\mu_{\alpha}(k)\rangle=E_{\alpha}(k)|\mu_\alpha(k)\rangle$, where $E_{\alpha}(k)$ ($|\mu_{\alpha}(k)\rangle$) denotes the eigenenergies (eigenstates) with $\alpha=\{-,0,+\}$ indexing the \{lowest, middle, highest\}-helicity branches, respectively.
Due to the $Z_2$ invariant, the system topology can be described by the Zak phase $\varphi_{\rm{Zak}}=\int^{\pi/d}_{-\pi/d}\langle\mu_-(k)| \partial_{k}|\mu_-(k)\rangle dk$ for the lowest branch.
The associated two distinct phases are characterized by the gauge-dependent Zak phase with $\varphi_{\rm{Zak}}=0$ or $\pi$ representing the topological trivial (or nontrivial) state.
We first ignore the gauge potential $\epsilon({k})$ which does not affect the topological invariant of the helicity branches and the topological phase transition of the system.
As a result, the Hamiltonian with satisfying the magnetic group symmetry ${\cal M}$ will ensure an inversion symmetry: ${\cal P}{\boldsymbol{h}}({k}){\cal P}={\boldsymbol{h}}({-k})$, where ${\cal P}= e^{-i\pi\hat{F}_z}$ is the inversion operator.
The Bloch states at the two higher symmetric momenta $\{k=0, k=\pi/d\}$ are eigenstates of ${\cal P}$: ${\cal P}|\mu_-(k=0)\rangle=P_1|\mu_-(k=0)\rangle$ and ${\cal P}|\mu_-k=\pi/d)\rangle=P_2|\mu_-(k=\pi/d)\rangle$.
Thus the winding number for the lowest helicity branch can be further experimentally extracted by measuring the $Z_2$ topological invariant $\nu=-{\rm {Im}}[{\rm {ln}}(P_1*P_2)]/\pi$~\cite{shorttopo}, which is equivalent to the Zak phase through $\nu=\varphi_{\rm{Zak}}/\pi$ with $\nu=1$ for topological nontrivial states and $\nu=0$ for trivial states. As to the experimental measurement, this property significantly improves the accuracy and accessibility of measurement the topological invariant in experiment~\cite{SCiGB2018TOPO}.
In addition, the spin texture for the higher symmetric momenta of the system can be extracted by the spin-resolved time-of-flight imaging~\cite{PhysRevA.94.061604}.

\begin{figure}[!htp]
\includegraphics[width=0.95\columnwidth]{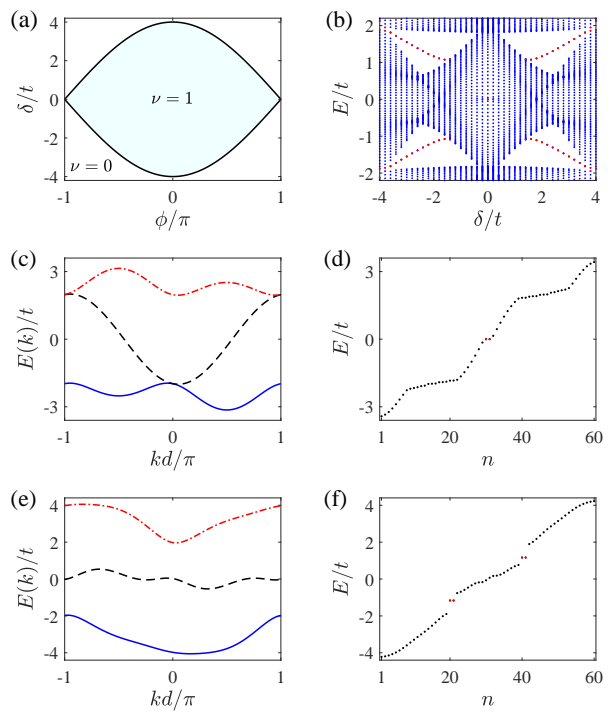}
\caption{\label{bandtopo}(color online). (a) Topological phase of Hamiltonian~(\ref{k-single}) on the $\phi$-$\delta$ plane. (b) The $\delta$-dependence energy spectra under OBC. Panels (c) and (e) shows the typical Bloch spectra for $\delta/t=0$ and $\delta/t=2$; the corresponding energy spectra under OBC are shown in (d) and (f). The red dots in (b), (d), and (f) denote the edge modes. The other parameters are chosen as $\phi/\pi=0.1$ and $t_y/t=1$.}
\end{figure}

Figure~\ref{bandtopo}(a) shows the phase diagram in the $(\phi, \delta)$-plane, where $\nu=1$ (0) correspond to topological nontrivial (trivial) phases for the lowest helicity branch.
The analytic $t_y$-independent phase boundary is given by $\delta/t=\pm4\cos(\phi/2)$.
As expected, the topological nontrivial phase $\nu=1$ exhibits topologically protected edge states due to the bulk-edge correspondence by imposing a hard-wall confinement along the $y$ axis; see Fig.~\ref{bandtopo}(b). We verify that the highest and lowest helicity branches possess identical topological invariant $\nu$, while the middle helicity branch is always topologically trivial with $\nu=0$.

Interestingly, the edge states under open boundary condition (OBC) depend on the strength of $\delta$.
In the absence of $\delta$, the system supports a dark state: $|\mu_0\rangle=[\mid\uparrow\rangle-\mid\downarrow\rangle]/\sqrt{2}$, in which the spin-$0$ component is decoupled from the spin-$\uparrow$ and $\downarrow$ components and the band spectra is gapless with touching points at higher symmetric momenta.
Figure~\ref{bandtopo}(d) displays the twofold degenerate zero-energy edge modes for $\delta=0$.
To understand this, we analyze the symmetries of Hamiltonian~({\ref{Eq.Ham 2}}).
In addition to the ${\cal M}$ symmetry, we find that the Hamiltonian~({\ref{Eq.Ham 2}}) satisfies $TH_0T^{-1}=-H_0$, where $T$ is the gauge transformation operator $\hat{a}^{\dag}_{n,\sigma} \longrightarrow (-1)^{n+1}\hat{a}^{\dag}_{n,\sigma}$.
 According to the definition, the two edge states $|\psi_L\rangle$ and $|\psi_R\rangle$ could transform into each other under the magnetic group symmetry: ${\cal M}|\psi_L\rangle=|\psi_R\rangle$ and ${\cal M}|\psi_R\rangle=|\psi_L\rangle$, which guarantees the twofold degeneracy. Moreover, the anticommutation relation between $H_0$ and $T$ implies that the two edge states have opposite energies.
Combining ${\cal M}$ and ${T}$ symmetries, the two zero-energy edge modes at left  ($|\psi_L\rangle$) and right ($|\psi_R\rangle$) are consistent with the numerical results shown in Fig.~\ref{bandtopo}(d). Therefore, the two zero-energy  edge states ($|\psi_L\rangle, |\psi_R\rangle$) are protected by both the ${\cal M}$ and ${T}$ symmetries. Remarkably, the $T$ symmetry forces the left edge state $|\psi_L\rangle$ only occupying the odd lattice site.

For nonzero values of $\delta$, there appear two pairs of edge states with nonzero opposite energies for a topological nontrivial phase [Fig.~\ref{bandtopo}(g)], which can also be well understood by analysing the system symmetry.
In contrast, to preserve $T$ symmetry at $\delta=0$, the Hamiltonian ({\ref{Eq.Ham 2}}) breaks $T$ symmetry when $\delta\neq0$.
However, the ${\cal M}$ symmetry ensures each pair of degenerate edge modes exhibiting nonzero energies, whereas the $\delta$ term fails to split out the degeneracy of each pair of edge modes.  Meanwhile, the system hosts a different gauge transformation symmetry $T'H_0T'^{-1}=-H_0$ with $T'=e^{-i\pi \hat{F}_{x}}\otimes T$. As a result, the commutation relation $[{\cal M },H_0]=0$ and the anticommutation relation $\{{T'},H_0\}=0$ support two pairs of edge states with opposite nonzero energies for topological nontrivial phases at nonzero $\delta$, leading to the four edge states as shown in Fig.~\ref{bandtopo}(f). Moreover, we find that these edge modes may occupy both even and odd lattice sites due to the broken of $T$ symmetry.

\section{Bulk-edge correspondence\label{Sec4}}
Now we turn to study the bulk-edge correspondence for the STMC spin-1 system including dissipation ${\boldsymbol{h}}({k})={\boldsymbol{h}}_0({k}) + i {\cal B}_x\hat{F}_x + i{\cal B}_z\hat{F}_z$. Here ${\cal B}_x$ (${\cal B}_z$) is the non-Hermitian parameters of in-plane (out-of-plane) imaginary magnetic fields~\cite{PhysRev.87.410}. As to the experimental feasibility, the imaginary magnetic field ${\cal B}_x$ and ${\cal B}_z$ can be realized by using a hyperfine resolved two-photon Raman process~\cite{PhysRevX.8.031079,Sci.Rep.10.1113}.
As a result, an effective non-Hermitian coupling term is emerged by adiabatically eliminating the excited state.

\begin{figure}[!htp]
\includegraphics[width=0.96\columnwidth]{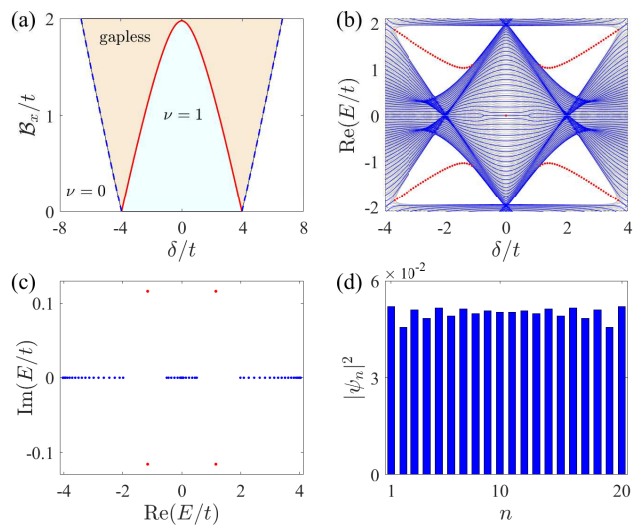}
\caption{\label{nonFx}(color online). (a) Phase diagram of non-Hermitian system in the $(\delta, {\cal B}_x)$ plane. (b)The real part of spectra as functions of $\delta$ with ${\cal B}_x/t=0.2$. The bulk spectra under PBC (gray lines) are the same as the ones based on OBC (blue lines), which demonstrates the bulk-edge correspondence. (c) The typical spectra under OBC for $\delta/t=2$. The red dots in (b) and (c) represent the edge states. (d) The typical density distribution of bulk state at $\delta/t=2$. The other parameters are $\phi/\pi=0.1$ and $t_y/t=1$.}
\end{figure}

To gain more insight, we first consider the atoms subjected to a ${\cal B}_x$ field.
The non-Hermitian term of $i{\cal B}_x \hat{F}_x$ preserves the ${\cal M}$ symmetry, which ensures that the topology of the non-Hermitian system remains characterized by the $Z_2$ invariant.
Figure~\ref{nonFx}(a) shows the phase diagram in the $(\delta, {\cal B}_x)$ plane with $\phi/\pi=0.1$.
When the bulk gap closes at the exceptional points, i.e., $E_{-}(k)=E_{0}(k)$, there exists two phase transition boundaries by tuning $\delta$. The red solid line denotes the phase transition between the gapped and gapless topological nontrivial phases,
while the topological trivial ($\nu=0$) to nontrivial ($\nu=1$) phase transition is characterized by the blue dashed line.
As can be seen, the region of topological nontrivial phase is roughly linear growing with ${\cal B}_x$.
Moreover, we find that the system exhibits a gapless phase in the topological nontrivial region, where the real part of the complex band gap closes but the imaginary part remains open.
The gapless phase region is largely enhanced at larger ${\cal B}_x$.

Figure~\ref{nonFx}(b) plots the real part of $E$ as a function of $\delta$ under PBC (gray lines), which is consistent with the bulk spectra under OBC (blue dots) excluding the expected two pairs edge states (red dots) for topological nontrivial phases.
The bulk-edge correspondence for our non-Hermitian spin-1 model is protected by the ${\cal M}$ symmetry, which guarantees the real Bloch energy spectra for the gapped phases with $E(k)=E^*({k})$ [Fig.~\ref{nonFx}(c)].
Figure~\ref{nonFx}(d) shows the density distribution of the bulk state along the lattice sites. As can be seen, the random bulk eigenstate for the gapped phase under OBC is a Bloch state due to the ${\cal M}$ protected bulk-edge correspondence.
To understand this, we assume an ansatz in which the bulk eigenstate with eigenenergy $E(k)$ takes the following form~\cite{PRLYS212018NT}:
$|\psi\rangle=|\mu(k)\rangle\bigotimes\sum_n(re^{ikd})^n|n\rangle$, where $r$ is real and positive decay index and $|n\rangle =[\hat{a}_{n,\uparrow},\hat{a}_{n,0}, \hat{a}_{n,\downarrow}]^T$ is the atomic state for $n$th lattice site.
Then we have ${\cal M}|\psi\rangle=[{\cal M}|\mu(k)\rangle]\bigotimes\sum_n r^{-n}e^{inkd}|n\rangle$ corresponding to the eigenenergy $E^*(k)$.
By utilizing the commutation relation $[{\cal M},{\boldsymbol{h}}({k})]=0$ and $|\mu(k)\rangle={\cal M}|\mu(k)\rangle$, which yields the decay index $r=1$ (extended state). Notably, the non-Hermiticity with preserving bulk-edge correspondence is very different from the ones in non-Hermitian parity-time symmetric systems~\cite{PhysRevLett.110.243902,PhysRevLett.115.200402,NatmatWei2017,Xiao2017,PhysRevB.84.153101}.

\begin{figure}[!htp]
\includegraphics[width=1\columnwidth]{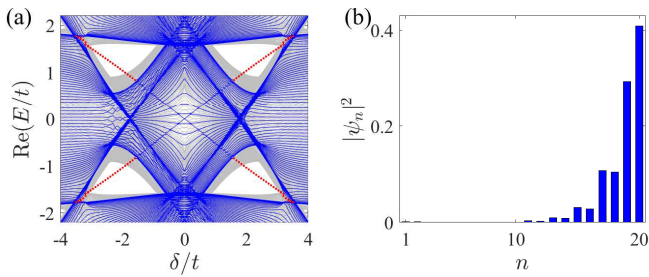}
\caption{\label{nonFz}(color online). (a) $\delta$ dependence of the real part of Bloch spectra. The gray lines denote the periodic Bloch spectra under PBC, while the blue lines (red dots) represent the bulk (edge) spectra under OBC. (b) The typical density distribution of bulk state for the topological nontrivial phase with $\delta/t=2$, which indicates the non-Hermitian skin effect with broken bulk-edge correspondence. The other parameters are $\phi/\pi=0.3$, $t_y/t=1$, and ${\cal B}_z/t=1.5$.}
\end{figure}

To further explore the bulk-edge correspondence in our high-spin system, we assure the system is illuminated by a ${\cal B}_z$ field.
Different from the $i{\cal B}_x \hat{F}_x$ term, the non-Hermitian term of $i{\cal B}_z \hat{F}_z$ breaks the ${\cal M}$ symmetry, which indicates that the complex band spectra could exist even for the gapped phases. Interestingly, we find that the real part of the Bloch band gap closes and reopens at $\delta/t=\pm4\cos(\phi/2)$, where the ${\cal B}_z$ independent phase boundary is the same as the Hermitian Hamiltonian~(\ref{k-single}). And the imaginary part of the Bloch band gap is equal to the value of ${\cal B}_z$ at the topological phase transition.

Figure~\ref{nonFz}(a) shows the real part of the energy spectra of the non-Hermitian system for $\phi/\pi=0.3$. Here, the gray lines correspond to the PBC, and the black and red dots respectively correspond to bulk and edge states under OBC. It's clear that the Bloch band spectra is quantitatively different from the open-boundary spectra. The significant divergence is ascribed to the broken of bulk-edge correspondence when ${\cal B}_z\neq0$ and $\phi\neq0$.
In particular, the non-Hermitian skin effect~\cite{PRLYS212018NT} is observed, as shown in Fig.~\ref{nonFz}(b), where the bulk eigenstate is localized near the boundary instead of the extended Bloch state. However, we verify that the non-Hermitian system in the presence of nonzero ${\cal B}_z$ can also host the bulk-edge correspondence when $\phi=0$. In this case, the bulk-edge correspondence is protected by the inversion symmetry (${\cal P}$) even in the absence of ${\cal M}$ symmetry.

\section{Conclusion\label{Sec5}}
Based upon the currently availably techniques for ultracold atoms, we propose the STMC spin-1 lattice model with tunable synthetic magnetic flux lattice, which possesses the SPT phase under the ${\cal M}$ symmetry.
Beyond the conventional AZ classification corresponding to the chiral symmetry protection, we explore the band topology and find two kinds of edge states under different symmetries.
Subsequently, we have investigated the bulk-edge correspondence in a non-Hermitian spin-1 model with including the imaginary Zeeman field.
In particular, under unbroken (broken) ${\cal M}$ symmetry, there appear preserved (destroyed) bulk-edge correspondence without (with) non-Hermitian skin effect.
Our study can be extended to study other spin-tensor-momentum coupled exotic topological quantum matters~\cite{PRLCZ2017STMC,PRLCZ2018STMC,PhysRevA.97.053609} and higher-order topological phase transitions~\cite{PRLTL2019NT,PhysRevLett.123.073601,PhysRevB.99.081302}.

\acknowledgements{This work is supported by the National Key R$\&$D Program of China (Grant No. 2018YFA0307500), the Key-Area Research and Development Program of GuangDong Province under Grant No. 2019B030330001, the NSFC (Grants No. 11874433, No. 11874434, and No. 11574405), and the Science and Technology Program of Guangzhou (China) under Grant No. 201904020024.}


%

\end{document}